\documentstyle[aps,epsfig]{revtex}

\begin{document}
\title{
Suppression of High-$P_T$ Jets as
a Signal for Large Extra Dimensions
and New Estimates of Lifetimes
for Meta stable Micro Black Holes\\
-From the Early Universe to
Future Colliders }

\author{Stefan Hofmann$^a$
, 
Marcus Bleicher$^b$
, 
Lars Gerland$^c$
, 
Sabine Hossenfelder$^a$
,
Sascha Schwabe$^a$
and Horst St\"ocker$^a$
}

\address{
$^a$ Institut f\"ur Theoretische Physik\\
J. W. Goethe Universit\"at\\
60054 Frankfurt am Main, Germany\\
$^b$ SUBATECH, Ecole des Mines\\
F-44307 Nantes, Cedex 03, France\\
$^c$ School of Physics and Astronomy\\
Tel Aviv University\\
69978 Tel Aviv, Israel 
}

\maketitle

\noindent
\begin{abstract}
We address the production of black holes at LHC
in space times with compactified space-like 
large extra dimensions (LXD).
Final state black hole production leads
to suppression of high-$P_T$ jets,
i.e. a sharp cut-off in
$\sigma$(pp$\rightarrow$jet$+X$).
This signal is compared to the jet plus missing
energy signature due to graviton
production in the final state
as proposed by the ATLAS collaboration.

Time evolution and lifetimes of the newly created
black holes are calculated based on the microcanonical
formalism. It is demonstrated that previous lifetime estimates of
micro black holes have been dramatically underestimated.

The creation of a large number of quasi-stable 
black holes is predicted
with life times of hundred fm/c at LHC.

Medium modifications of the black holes evaporation rate
due to the quark gluon plasma in relativistic 
heavy ion collisions as well as 
provided by the cosmic fluid in the early universe are studied.
\vspace{1cm}
\end{abstract}

An outstanding problem in physics is to understand the ratio
between the electroweak scale $m_W=10^3$~GeV
and the four-dimensional Planck scale $m_{\rm P}=10^{19}$~GeV.
Proposals that address this so called
hierarchy problem within the context of brane world scenarios
have emerged recently \cite{add}.
In these scenarios the Standard Model of particle physics
is localised on a three dimensional brane in a higher
dimensional space with large compactified
space-like extra dimensions (LXD). 
This raises the exciting possibility
that the fundamental scale $M_f$ can be as low as $m_W$
(without extra dimensions $M_f=m_{\rm P}$.
As a consequence, future high energy colliders like LHC
and CLIC
could probe the scale of quantum gravity with its exciting new 
phenomena: A possible end of small
distance physics has been investigated by Giddings\cite{gid} 
while Dimopoulos 
and Landsberg speculated on the production of black holes in high 
energetic interactions \cite{dim}.
In this letter we investigate TeV scale gravity 
associated with black hole production and evaporation
at colliders and in the early universe.

One scenario for realizing TeV scale gravity is a
brane world in which the Standard Model particles
including gauge degrees of freedom reside on a 
3-brane within a flat compact space of volume
$V_d$, where $d$ is the number of LXDs with radius $L$.
Gravity propagates in both the compact LXDs and non-compact
dimensions.

Let us first characterise black holes
in space times with LXDs.
We can consider two cases.
First, the size of the black hole given by
its Schwarzschildradius is $R_H \gg L$.
In this case the topology of the horizon
is $S(3)\times U(1) \times U(1)\times \dots$,
where $S(3)$ denotes the three dimensional sphere
and $U(1)$ the Kaluza-Klein compactification.
Second, if $R_H \ll L$ the topology of the horizon
is spherical in $3+d$ space-like dimensions.

The mass of a black hole with $R_H \approx L$
in $D=4$ is called the critical mass
$M_c\approx m_{\rm P} L/l_{\rm P}$ and $1/l_{\rm P}=m_{\rm P}$.
Since 
$ 
L\approx
\left(1 {\rm TeV}/M_f \right)^{1+\frac{2}{d}}
10^{\frac{31}{d}-16} \; {\rm mm}
$,
$M_c$ is typically of the order of the Earth mass. 
As we are interested in black
holes produced in parton-parton collisions
with a maximum cms energy of $\sqrt{s}=14$~TeV,
these black holes have $R_H \ll L$ 
and belong to the second case.

Spherically symmetric solutions describing black holes
in $D=4+d$ dimensions have been obtained\cite{hor}
by making the ansatz
\begin{equation}
{\rm d}s^2 
=
- {\rm e}^{2\phi(r)} {\rm d}t^2
+ {\rm e}^{2\Lambda(r)} {\rm d}r^2
+ r^2 {\rm d}\Omega_{(2+d)}
\; ,
\end{equation}
with ${\rm d}\Omega_{(2+d)}$ denoting the
surface element of a unit ($3+d$)-sphere.
Solving the field equations $R_{\mu\nu} = 0$
gives
\begin{equation}
{\rm e}^{2\phi(r)} = {\rm e}^{-2\Lambda(r)}
=
1-\left(\frac{C}{r}\right)^{1+d},
\end{equation}
with $C$ being a constant of integration.  
We identify $C$ by the requirement
that for $r\gg L$ the force derived from 
\begin{equation}
V(r)
=
\frac{1}{1+d} 
\frac{1}{M_f^{1+d}} \frac{M}{M_f} \; \frac{1}{L^d} \;
\frac{1}{r}
\end{equation}
in a space-time with $d$ compactified extra dimensions 
equals the $4$-dimensional Newton force.
Note, the mass $M$ of the black hole is 
\begin{equation}
M
\approx
\int {\rm d}^{3+d} x
\; T_{00}
\end{equation}
with $T_{\mu\nu}$ denoting the energy momentum tensor
which acts as a source term in the Poisson equation
for a slightly perturbed metric in
$3+d$ dimensional space-time \cite{my}.
Note that in order to fix partially the coordinate
invariance the harmonic gauge condition is imposed.
In this way the horizon radius is obtained as
\begin{equation}
R_H^{1+d}=
\frac{2}{1+d} \; 
\left(\frac{1}{M_f}\right)^{1+d} \; \frac{M}{M_f}
\end{equation}
with $M$ denoting the black hole mass.

Let us now investigate the production rate
of these black holes at LHC. 
Consider two partons moving in opposite
directions. If the partons cms energy
$\sqrt{\hat s}$ reaches the fundamental
scale $M_f\sim 1$~TeV
and if the impact parameter is less than $R_H$,
a black hole with Mass $M\approx \sqrt{\hat s}$
might be produced.
The total cross section for such a process
can be estimated on geometrical grounds
and is of order $\sigma(M)\approx \pi R_H^2$.
This expression contains only the fundamental
scale $M_f$ as a coupling constant.
As a consequence, if we set $M_f\sim 1$TeV and
$d=2$ we find 
$\sigma \approx \pi$~TeV$^{-2}\approx 1.2$~nb.
However, we have to take into account
that in a pp-collision 
each parton carries only a fraction of the
total cms energy. The relevant quantity
is therefore 
the Feynman $x$ distribution
of black holes at LHC for masses
$M\in[M^{-},M^{+}]$ 
given by
\begin{equation}
\frac{d\sigma}{d x_F} = \sum\limits_{p_1,p_2} 
\int\limits_{M^{-}}^{M^{+}} dy 
\frac{2 y}{x_2 s} f_1(x_1, Q^2) f_2(x_2, Q^2) 
\sigma(y,d)\; ,
\end{equation}
with $x_F=x_2-x_1$ and the restriction 
$x_1 x_2 s=M^2$.
We used the CTEQ4 \cite{lai1} parton distribution functions $f_1$, $f_2$ 
with $Q^2=M^2$. 
All kinematic combinations of partons from 
projectile $p_1$ and target $p_2$ are summed over. 
\begin{figure}[h]
\vskip 0mm
\vspace{0cm}
\centerline{\psfig{figure=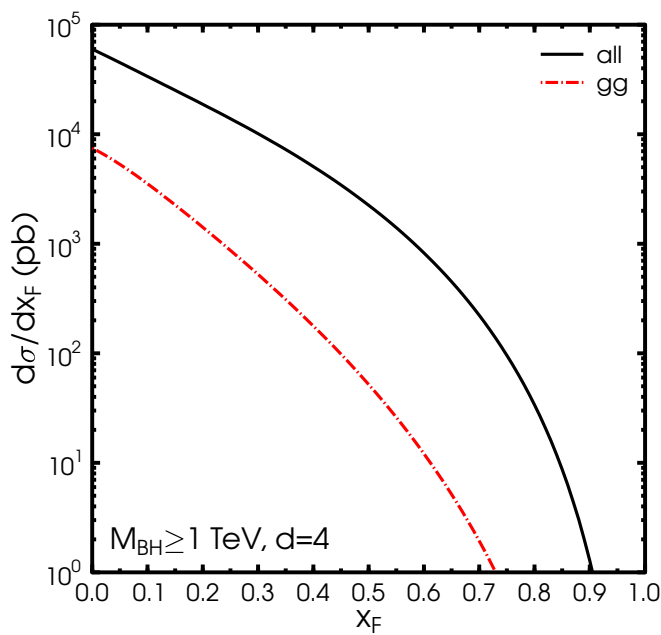,width=3.5in}}
\vskip 2mm
\caption{Feynman $x$ distribution of black holes  
with $M\ge 1$~TeV produced in pp interactions at LHC with 4 
compactified spatial extra dimensions 
\protect\cite{we0}.
\label{dndy}}
\end{figure}
Fig. \ref{dndy} depicts the momentum distribution of produced 
black holes in pp interactions at $\sqrt s = 14$~TeV.
Since for masses below $10$~TeV
heavy quarks give a vanishing contribution 
to the black hole production cross section,
those black holes are primarily formed in
scattering processes of
valence quarks. 

The possibility of producing black holes
in pp collisions at LHC has dramatic consequences.
First of all this would result in a sharp cut-off
in $\sigma$(pp$\rightarrow$jet$+X$) as a
function of the transverse jet energy at $E_T\sim M_f$.
A different extra dimension signature
occurring in the same process was discussed
in an ATLAS proposal \cite{ATLAS}. The authors of this
proposal found for specific model parameters
a significant decrease in $\sigma$(pp$\rightarrow$jet$+X$)
at $E_T \sim M_f$ due to gravitons $G$ in the final states.
This effect is somehow contraintuitive since the
coupling of any Kaluza-Klein state to our brane
is still suppressed by $1/m_{\rm P}^2$.
For instance, the total cross section \cite{Ruba}
for the subprocess $q\bar{q}\rightarrow \gamma G$ 
is of order $\sigma$($q\bar{q}\rightarrow \gamma G$)
$\sim \alpha_s/m_{\rm P}^2 N(\sqrt{\hat{s}})$,
where $N$ denotes the number of Kaluza-Klein states
with masses below $\sqrt{\hat{s}}$.
Since the momenta of the Kaluza-Klein states 
perpendicular to our brane are quantised
in integer multiples of $1/L$ and 
the norm of these momenta are the four-dimensional
masses of the Kaluza-Klein states, one has
$N(\sqrt{\hat{s}})\approx L\sqrt{\hat{s}}/2\pi$
for each extra dimension. As a result,
$\sigma$($q\bar{q}\rightarrow \gamma G$)
$\sim \alpha_s/M_f^2 (\sqrt{\hat{s}}/M_f)^d$.
Hence, the total cross section rapidly
increases with the partons cms energy
and for $\sqrt{\hat{s}} \sim M_f$ 
it is comparable to parton parton cross sections in pQCD.
However, this jet plus missing transverse energy
signature strongly depends on the model parameters
and is for $M_f>5$TeV extremely difficult to measure
in the minimal scenario with $d=2$.
In contrast, a small scale cut-off due to
black holes in the final state would be 
a strong signature. One might argue that 
for $M_f>5$TeV this signature would be
difficult to measure, too. 
This is not necessarily true, since a small scale cut-off
corresponds to a black hole 
in the final state, which radiates off 
quanta at Temperatures $T_H\sim M_f (M_f/M)^{1/(d+1)}$.
The process of evaporation can result in
a bump of the standard cross section for 
jet production at much lower energies $E\sim T_H$.
The previous discussion will be given in
detail in a forthcoming publication \cite{we1}.

Let us now investigate the evaporation of 
black holes with $R_H \ll L$ and 
study the influence of compact extra dimensions 
on the emitted quanta.
In the framework of black hole thermodynamics 
the entropy $S$ of a black hole is given by its surface
area. In the case under consideration
$S\sim\Omega_{(2+d)} \; R_H^{2+d}$.
The single particle spectrum 
of the emitted
quanta in the microcanonical ensemble 
is then \cite{Harms}
\begin{equation}
n(\omega) 
= 
\frac{{\rm exp}[S(M-\omega)]}{{\rm exp}[S(M)]}\quad.
\end{equation}
It has been claimed that it may not be possible
to observe the emission spectrum directly,
since most of the energy is radiated in
Kaluza-Klein modes. However, from the
higher dimensional perspective this seems
to be incorrect and most of the energy
goes into modes on the brane.

Summing over all possible multi particle spectra 
we obtain the BH's evaporation 
rate $\dot{M}$ through the 
Schwarzschild surface ${\cal A}_D$ in $D$ space-time dimensions,
\begin{equation}
\dot{M} = - {\cal A}_{D} \; \frac{\Omega_{(2+d)}}{(2\pi)^{3+d}} \; 
\int\limits_0^{M} 
{\rm d}\omega\; \sum\limits_{j=1}^{(M/\omega)} 
\omega^{D-1} n(j\omega) \quad.
\label{upsmdot}
\end{equation}
Neglecting finite size effects 
eq.(\ref{upsmdot}) becomes
\begin{equation}
\dot{M} = {\cal A}_D
\; \frac{\Omega_{(2+d)}}{(2\pi)^{3+d}} \; 
{\rm e}^{-S(M)} \sum\limits_{j=1}^{\infty}
\left(\frac{1}{j}\right)^D 
 \int\limits_{M}^{(1-j)M} {\rm d}x\; 
(M-x)^{D-1} {\rm e}^{S(x)} \Theta(x)\quad, 
\end{equation}
with $x= M- j\omega$, 
denoting the energy of the black hole after emitting $j$ 
quanta of energy $\omega$. 
Thus, ignoring finite size effects 
we are lead to the 
interpretation that the black hole emits only a single 
quanta per energy interval.
We finally arrive at
\begin{equation}
\label{mdoteq}
\dot{M} = {\cal A}_D 
\; \frac{\Omega_{(2+d)}}{(2\pi)^{3+d}} \; 
\zeta (D) 
\int\limits_0^M {\rm d} x \; 
(M-x)^{D-1} {\rm e}^{S(x)-S(M)}
\; .
\end{equation}
\begin{figure}[h]
\vskip 0mm
\vspace{0cm}
\centerline{\psfig{figure=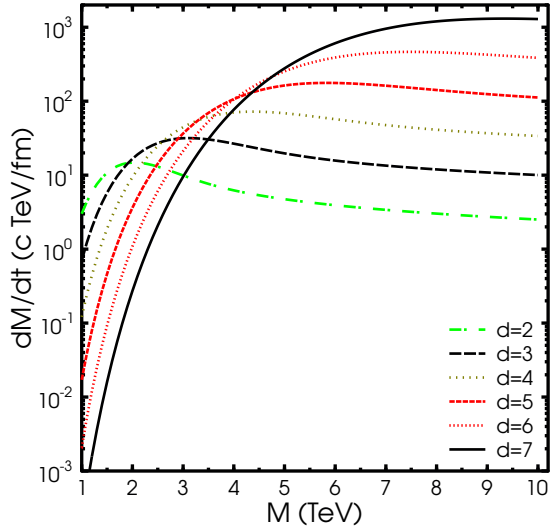,width=3.5in}}
\vskip 2mm
\caption{Decay rate in TeV/fm as a function of the 
initial mass of the black hole \protect\cite{we0}. 
Different line styles correspond to different numbers of 
extra dimensions $d$.  
\label{mdot}}
\end{figure}
Fig. \ref{mdot} shows the decay rate (\ref{mdoteq})
in TeV/fm as a 
function of the initial mass of the black hole. 
Since the Temperature $T_H$ of the black hole 
decreases like $M^{-1/(1+d)}$ it is evident
that extra dimensions help stabilising
the black hole, too.
\begin{figure}[h]
\vskip 0mm
\vspace{0cm}
\centerline{\psfig{figure=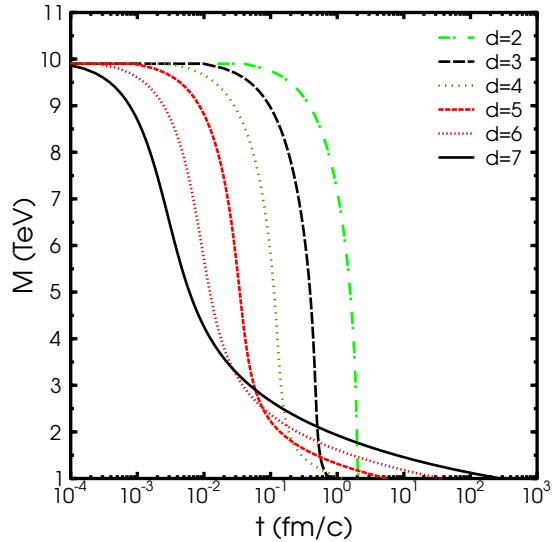,width=3.5in}}
\vskip 2mm
\caption{Time evolution of a black hole \protect\cite{we0}. 
Different line styles correspond to different numbers of 
extra dimensions $d$.  
\label{mt2}}
\end{figure}
From (\ref{mdoteq}) we calculate the time evolution
of a black hole with given Mass $M$. The result is 
depicted in Fig \ref{mt2} for different numbers
of compactified space-like extra dimensions.
As can be seen again, extra dimensions lead to an increase 
in lifetime of black holes. 
The calculation shows that a black hole with $M\sim$~TeV 
exists for $\sim 100$~fm/c (for $d>5$). Afterwards
the mass of the black hole drops below the fundamental
scale $M_f$. The quantum physics at this scale
is unknown and therefore the fate of the 
extended black object. 
However, statistical mechanics may still be valid.
If this would be the case it seems that after
dropping below $M_f$ a quasi-stable remnant remains.
%
%

Therefore, it is interesting to study the in
medium modification of the BH's evaporation rate (11).
In general the evaporation rate becomes
$$
\dot{M}
=
\dot{M}_{\rm loss} + \dot{M}_{\rm gain}
$$
The loss term describes the evaporation rate
which receives contributions from
the production of Kaluza-Klein states and
Standard Model particles. The Kaluza-Klein
states will be radiated into the $(3+d)$-dimensional space 
while Standard Model particles
are produced on our brane only.
In the following we assume that most
of the emitted quanta will be localised 
on our 3-brane.
The gain term takes into account the absorption
of medium particles on our brane.

An illustrative application is a medium
with a homogeneous energy density $\rho$
on our brane
which is constant in time.
The evaporation rate vanishes for
\begin{equation}
\rho
\approx
10^4 \left(\frac{M_f}{{\rm GeV}}\right)^2
\left(\frac{M_f}{M}\right)^{\frac{2}{1+d}}
\; \frac{\rm GeV}{{\rm fm}^3}
\; ,
\end{equation}
which is $10^9$ times the energy density
of the quark gluon plasma for
$M_f=1$TeV and $M\approx 10 M_f$.
In turn, if the density of the medium
corresponds to the energy density
of the quark gluon plasma,
the evaporation rate vanishes for
an initial mass of the black hole
of $M\approx 10^{18}$GeV.

One particular interesting scenario is the
evaporation of black holes in the early universe.
The medium is then provided by the total cosmic
fluid density $\rho$. The standard cosmological
evolution is rediscovered from the brane world
scenario whenever $\rho/m_{\rm P}^2$ is
large compared to $(M_f^{2+d}/m_{\rm P}^2)^{(2/d)}$,
so that the early time cosmology is analogous
to standard cosmology. In this early phase
the cosmic evolution is given by
the standard Friedmann equation in $D=4$,
$H^2\sim \rho/3m_{\rm P}^2$.
Thus, 
\begin{equation}
\dot{M}_{\rm gain}
\approx
\left(\frac{M}{M_f}\right)^{\frac{2}{1+d}} 
\left(\frac{m_{\rm P}}{M_f}\right)^2
H^2
\; .
\end{equation}
The black hole stops evaporating due to
the gain of energy from in-falling medium particles.
For $T>1$TeV this happens when the following
relation for the mass of the black hole and
the temperature of the cosmic fluid holds:
\begin{equation}
\frac{M(t)}{M_f}
\approx
\frac{10}{g_*(t)} 
\left(\frac{M_f}{{\rm GeV}}\right)^{1+d}
\left(\frac{{\rm GeV}}{T(t)}\right)^{2+2d}
\; ,
\end{equation}
where $g_*$ counts the effective degrees of
freedom of all relativistic particles.
The relevance of in-medium modifications
of the evaporation process for early universe
physics will be explored in a forthcoming
publication \cite{we2} in more detail.

In conclusion, we have predicted the momentum distribution
of black holes in space-times with LXDs.
We discussed the high $P_T$ suppression of jets
due to the formation of black holes in the
final state as a clear signature
for extra dimensions. We compared this observable
to the jet plus missing energy signature proposed
by ATLAS. The high $P_T$ suppression shows up
as a sharp cut-off in $\sigma$(pp$\rightarrow$jet$+X$)
in contrast to a rather smooth decrease caused
by gravitons in the final state.
Using the micro canonical ensemble
we calculated the decay rate of black holes
neglecting finite size effects.
If statistical mechanics is still valid below
the fundamental scale $M_f$, the black holes
may be quasi-stable. In the scenarios
($M_f\sim$~TeV, $d>5$) the lifetime is of order
$100$~fm/c.   
This motivated the study of in-medium modifications
of the black hole evaporation rate. We estimated
the evaporation rate for a black hole surrounded
by a homogeneous energy density which is constant
in time and for a black hole in the early universe.

This work was supported in parts by BMBF, DFG, and GSI.

\end{document}